\documentclass{aastex}      
\usepackage[dvips]{hyperref}
\usepackage{spr-astr-addons}

\newcommand{\gradad}{\nabla_{\mbox{\scriptsize adia}}}
\newcommand{\Ra}{\text{Ra}}
\newcommand{\KT}{{\cal K}_T}
\newcommand{\Amp}{{\cal A}}
\newcommand{\Fbot}{F_{\hbox{\scriptsize bot}}}
\newcommand{\Tbump}{T_{\hbox{\scriptsize bump}}}
\newcommand{\Tsurf}{T_{\hbox{\scriptsize surf}}}
\newcommand{\Kmin}{K_{\hbox{\scriptsize min}}}
\newcommand{\Kmax}{K_{\hbox{\scriptsize max}}}
\def\Div{\mathop{\hbox{div}}\nolimits}
\renewcommand{\na}{\vec{\nabla}}

\begin{document}
\title{Numerical simulations of the $\kappa$-mechanism with convection}


\author{T. Gastine\altaffilmark{1}} 
\and 
\author{B. Dintrans\altaffilmark{1}}
\email{thomas.gastine@ast.obs-mip.fr} 

\altaffiltext{1}{LATT, Universit\'e de Toulouse, CNRS, 14 avenue Edouard
Belin, F-31400 Toulouse, France}

\begin{abstract}
A strong coupling between convection and pulsations is known to play
a major role in the disappearance of unstable modes close to the red
edge of the classical Cepheid instability strip. As mean-field models
of time-dependent convection rely on weakly-constrained parameters  
\citep[e.g.][]{Baker},
we tackle this problem by the means of 2-D Direct Numerical
Simulations (DNS) of $\kappa$-mechanism with convection.

Using a linear stability analysis, we first determine the physical
conditions favourable to the $\kappa$-mechanism to occur inside a
purely-radiative layer. Both the instability strips and the nonlinear
saturation of unstable modes are then confirmed by the corresponding
DNS. We next present the new simulations with convection, where a
convective zone and the driving region overlap. The coupling between
the convective motions and acoustic modes is then addressed by using
projections onto an acoustic subspace.

\end{abstract}


\section{Introduction}

\cite{Eddington1917} discovered an excitation mechanism of stellar
oscillations that is related to the opacity behaviour in ionisation zones:
the $\kappa$-mechanism, where $\kappa$ denotes the opacity. This mechanism 
occurs when the opacity varies during compression phases so as to block the 
emerging radiative
flux \citep{zhevakin,cox58}. Ionisation
regions correspond to a strong increase in opacity that leads to the
\emph{opacity bumps} responsible for the local excitation of
modes. These driving zones should nevertheless be located at
a precise radius inside the star, neither too close to the surface
nor to deep, in order to balance the damping
that mainly occurs at the surface. It defines the so-called \emph{transition
region} that shapes the limit between the quasi-adiabatic interior and
the strongly non-adiabatic surface. For classical Cepheids that
pulsate in the fundamental acoustic mode, this transition region
is located at a temperature $T\simeq 4\times 10^4$ K corresponding to
the second helium ionisation \citep{Baker65}.  However,
the bump location is not solely responsible for the acoustic
instability. A careful treatment of the $\kappa$-mechanism would
involve dynamical couplings with convection, metallicity effects and
both
realistic equations of state and opacity tables \citep{bono99}. 
The purpose of our model is to simplify the hydrodynamic
approach while retaining the \emph{leading-order} phenomenon --e.g. the
location of the opacity bump or its amplitude-- such that feasible
direct numerical simulations of the $\kappa$-mechanism with convection
can be achieved.

\section{The first step: radiative models in 1-D}

We first focus on radial modes propagating in a purely-radiative and
partially-ionised layer and then restrict our study to the 1-D case. Our
model consists of a \emph{local zoom} about an ionisation region and
is composed by a monatomic and perfect gas (with $\gamma=c_p/c_v=5/3$)
under both a constant gravity $\vec{g}$ and kinematic viscosity
$\nu$. All quantities in our model are dimensionless, e.g.
temperature is given in unit of the surface temperature of the star,
length in unit of a fraction of the star radius and velocity in unit of
the sound speed at the surface.

\begin{figure}[t]
\centering
\includegraphics[width=8cm]{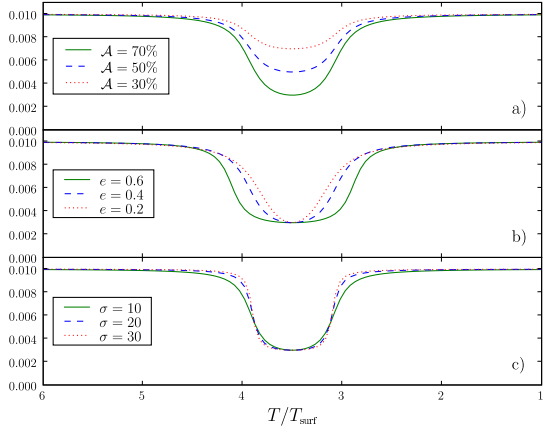}
\caption{Influence of the hollow parameters on the conductivity profile for
$\Kmax=10^{-2}$ and $\Tbump/\Tsurf=3.5$.}
\label{fig:profile-rad}
\end{figure}

In order to easily investigate the influence of the opacity bump on the
stability, we adopt the following conductivity hollow to mimic this bump
\cite[as $\kappa \propto 1/K$, see][hereafter Paper I]{paperI}:

\begin{equation}
 K_0(T_0)=\Kmax\left[1+\Amp\dfrac{-\pi/2+\arctan(\sigma
T^+T^-)}{\pi/2+\arctan(\sigma e^2)}\right],
\end{equation}
with $\Amp=(\Kmax-\Kmin)/\Kmax$ and $T^{\pm}=T_0-\Tbump \pm e$. Here
$T_0$ is the equilibrium temperature profile, $\Tbump$ is the hollow
central temperature, and $\sigma,\ e$ and $\Amp$ denote its slope,
width and relative amplitude, respectively. Examples of common values
of these parameters are provided in Fig.~\ref{fig:profile-rad}.

\subsection{Conditions for the instability}

Using the well-known work integral formalism \cite[e.g.][]{Unno-book},
it is easy to show that the following condition is necessary to get
unstable modes by $\kappa$-mechanism in variable stars:

\begin{equation}
 \dfrac{d\KT}{dz} < 0 \quad\text{where}\quad \KT\equiv
 \left(\dfrac{\partial \ln K_0}{\partial \ln T_0}\right)_{\rho_0}.
\label{eq:KT} 
\end{equation} 
However, this condition is not sufficient as the ionisation region
should
also be located at a given location in the star, neither too deep nor
too close to its surface. This area is the so-called \emph{transition
region} that separates the quasi-adiabatic interior from the strongly
non-adiabatic surface and is defined by
\cite[e.g.][]{Cox80}

\begin{equation}
 \Psi = \dfrac{\langle c_v T_0 \rangle \Delta m}{\Pi L} \simeq 1,
 \label{eq:psi}
\end{equation}
where $\Delta m$ is the integrated mass between the considered point and
the surface, $\Pi$ the mode period and $L$ the luminosity. $\Psi$
is the ratio between the thermal energy embedded between
the given radius and the surface and the energy radiated during an
oscillation period.

\begin{figure}[t]
 \centering
 \includegraphics[width=8cm]{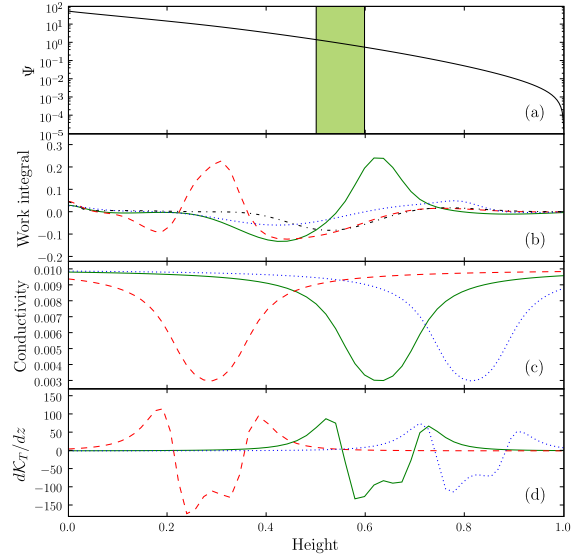}
 \caption{\textbf{a}) Vertical profile of the $\Psi$ coefficient
 (Eq.~\ref{eq:psi}). The superimposed green zone represents
 the location where $\Psi = 1\pm 0.5$. \textbf{b}) The real part of
 the work integral plotted for the three different equilibrium models. The
 dot-dashed black line is for the constant radiative conductivity case.
 \textbf{c}) Corresponding radiative conductivity
 profiles. \textbf{d}) Corresponding equilibrium fields $d \KT /dz$
(Eq.~\ref{eq:KT}).}
 \label{fig:wi}
\end{figure}

\subsection{The linear stability analysis}

In order to check these two conditions, we perform a linear stability
analysis by using the Linear Solver Builder code
\citep{LSB}. By starting with the conductivity profiles shown in
Fig.~\ref{fig:profile-rad}, we first compute the equilibrium setup that
is solution of both the hydrostatic and radiative equilibria given by

\begin{equation}
\vec{\nabla} p_0 = \rho_0 \vec{g}\quad \text{and}\quad \Div
[K_0(T_0)\vec{\nabla}T_0] = 0,
\label{eq:equil}
\end{equation}
where $p_0$ and $\rho_0$ denote the equilibrium pressure and
density. Then, we solve the linear oscillation equations for the
perturbations by seeking normal modes of the form $\exp(\lambda t)$
with $\lambda=\tau+i\omega$, where $\omega$ is the mode frequency and
$\tau$ its growth or damping rate (unstable modes corresponding to $\tau
> 0$). By varying the hollow parameters, we then determine
the physical conditions that lead to unstable modes excited by the
$\kappa$-mechanism in our layer (cf. Paper I).

The two criteria (\ref{eq:KT}) and (\ref{eq:psi}) predict that 
modes are unstable for a ``sufficient'' hollow (adequate
amplitude and width) that is located in the transition region. Using
the work integral computed from the obtained eigenvectors, we indeed
recover instability strips for the $\kappa$-mechanism by considering
the three different equilibrium setups (Fig.~\ref{fig:wi}):

\begin{itemize}
\item The first case (dotted blue line, $\Tbump/\Tsurf=1.7$) corresponds to a
hot star with an
ionisation region close to the surface. As the 
surrounding density
is small, $\Psi \ll 1$ and the conductivity
hollow hardly influences the work integral: driving is
unable to prevail over damping, leading to $\tau < 0$ or \emph{stable} modes.

\item In the second case (solid green line, $\Tbump/\Tsurf=2.1$), the radiative
conductivity
begins to decrease significantly at the location of the transition
region where $\Psi \simeq 1$. As a consequence, driving becomes
important. Furthermore, the radiative conductivity increase occurs where
non-adiabatic effects are already significant, i.e. $\Psi < 1$ there. It means
that no damping occurs between the hollow position and the surface as
the radiative flux perturbations are frozen-in. Driving is overcoming
damping in this case, therefore $\tau > 0$ and modes are \emph{unstable}.

\item The third case (dashed red line, $\Tbump/\Tsurf=2.8$) corresponds to a
cold star of which
ionisation region is located deeper in the stellar atmosphere where $\Psi
\gg 1$. Ionisation then occurs in a quasi-adiabatic location. As a
consequence, the excitation provided by the conductivity hollow cannot
balance the damping arising at the top of the layer, thus $\tau <0$ and 
modes are
\emph{stable}.

\end{itemize}

In conclusion, our first radiative model of the $\kappa$-mechanism is
well able to reproduce the main physics involved in the oscillations of
Cepheid stars: \textit{(i)} one should have $d \KT /dz <0$ to drive the
oscillations; \textit{(ii)} the thermal engine underlying the conductivity
hollow should also be located in the transition region where $\Psi \simeq
1$ (Fig.~\ref{fig:wi}).

\subsection{Study of the nonlinear saturation}

To confirm both the results obtained previously in the linear stability
analysis and to investigate the resulting nonlinear saturation, we perform
direct numerical simulations of our problem. The idea is to start from
the most favourable initial conditions found in the stability analysis and
then advance the general hydrodynamic equations in time. We use a
modified version of the public-domain finite-difference Pencil Code\footnote[2]{See
\url{http://pencil-code.googlecode.com/}.} that includes an implicit
solver of our own for the radiative diffusion term in the temperature
equation \cite[][hereafter Paper II]{paperII}.

\begin{figure}[t]
 \centering
 \includegraphics[width=8cm]{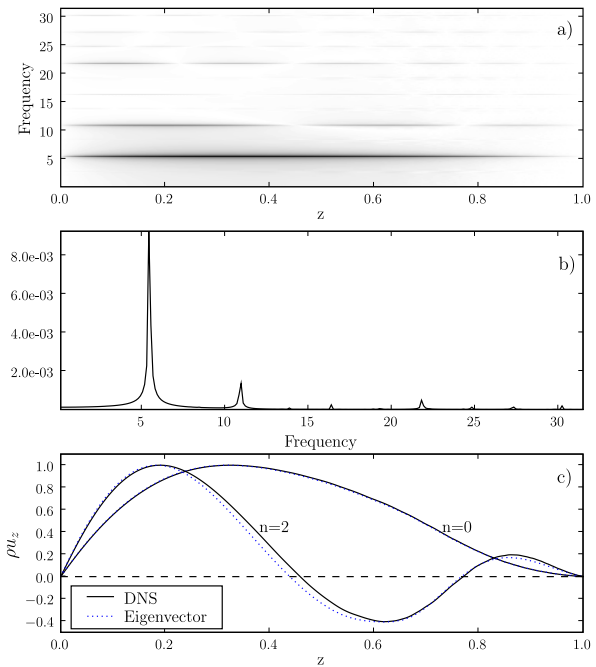}
\caption{\textbf{a)} Temporal power spectrum for the momentum in the
$(z,\ \omega)$-plane. \textbf{b)} The resulting spectrum after
integrating over depth. \textbf{c)} Comparison between normalised momentum
profiles for $n=(0,\ 2)$ modes according to the DNS power
spectrum (solid black line) and to the linear stability analysis
(dotted blue line).}
 \label{fig:fourier}
\end{figure}

To determine which modes are present in the DNS in the
nonlinear-saturation regime, we first perform a temporal Fourier transform
of the momentum field $\rho u(z,t)$ and plot the resulting power spectrum
in the $(z,\omega)$-plane (Fig.~\ref{fig:fourier}a). The mean spectrum
follows by integrating $\widehat{\rho u}(z,\omega)$ over depth
(Fig.~\ref{fig:fourier}b). 
With this method,
acoustic modes are extracted because they emerge as ``shark-fin profiles''
about definite eigenfrequencies \citep{dintrans04}. 

Several discrete peaks corresponding to
normal modes well appear but the fundamental mode close to $\omega_0=5.439$
clearly dominates. Moreover, the linear eigenfunctions are compared
to the mean profiles computed from a zoom taken in the DNS power
spectrum about eigenfrequencies $\omega_0=5.439$ and $\omega_2=11.06$
(Fig.~\ref{fig:fourier}c). The agreement between the linear
eigenfunctions (dotted blue lines) and the DNS profiles
(solid black lines) is remarkable. In summary, Fig.~\ref{fig:fourier}
shows that several overtones are present in the DNS, even for long times.
These overtones are however \emph{linearly stable} suggesting that 
some underlying
energy transfers occur between modes through nonlinear couplings.

These nonlinear interactions are investigated thanks to a tool
that measures the sound generation by turbulent flows
\cite[e.g.][]{Bogdan}. It involves projections of the DNS physical fields
onto the regular and adjoint eigenvectors that are
solutions to the linear oscillation equations. The projection
coefficients then give 
the time evolution of each
acoustic mode that is present in the DNS \emph{separately}. Furthermore,
as the evolution of the kinetic energy of each mode is also accessible with
this method, one can follow the energy transfer between modes.

\begin{figure}[h]
 \centering
 \includegraphics[width=8cm]{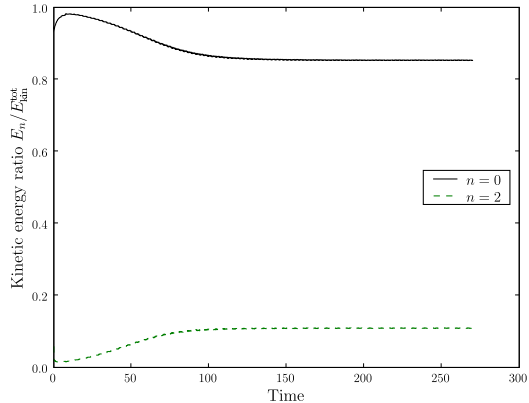}
 \caption{Time evolution of the kinetic energy ratio for the $n=0$
and $n=2$ acoustic modes.}
\label{fig:nrj}
\end{figure}

Figure \ref{fig:nrj} shows an example of the time evolution of the
kinetic energy ratio $E_n/E^\text{tot}_\text{kin}$ for the two modes 
$n = 0$ (i.e. the fundamental one) and $n=2$ (i.e. the second overtone). Here
$E_n$ denotes the kinetic energy embedded in the acoustic mode of order
$n$, while $E^\text{tot}_\text{kin}$ is the total kinetic energy in
the simulation.
After its linear transient growth,
a given fraction of the fundamental mode energy is progressively transferred to
the second overtone until the nonlinear saturation is achieved above time
$t\simeq 150$, with finally $85\%$ of the total kinetic energy 
in the $n=0$ mode and the remaining $15\%$ in the $n=2$ one. We recall
that it is at a first glance surprising to detect this second overtone
at long times because it is linearly stable and should normally be
damped by diffusive effects.  

The reason for its presence lies in a favored coupling
in the period ratio between these two modes. Indeed, the
fundamental period is $P_0=2\pi/\omega_0\simeq 1.155$ in this simulation 
while the
$n=2$ one is $P_2\simeq 0.568$. As a consequence, the corresponding period
ratio is close to one half ($P_2/P_0\simeq 0.491$) such that the second
overtone is
involved in the nonlinear saturation
through a 2:1 resonance with the fundamental mode.
The linear growth of the fundamental mode is then balanced by the 
pumping
of energy to the stable second overtone that behaves
as an energy sink, leading to the full limit-cycle stability.

\section{Convective models in 2-D}

\subsection{From radiative to convective setups}

After the previous results obtained on purely radiative models in 1-D, we
next address the convection-pulsation coupling that is suspected to
quench the radial oscillations of Cepheids close to the red edge.
The idea is to slightly modify the unstable 
radiative setup to obtain a convective zone superimposed to the ionisation
region in 2-D simulations. The convective instability obeys Schwarzschild's
criterion given by \cite[e.g.][]{Chandra61}

\begin{equation}
 \left| \dfrac{dT_0}{dz}\right| > \dfrac{g}{c_p} \quad\Longrightarrow\quad
\Fbot > \dfrac{g K_0(T_0)}{c_p}.
\label{eq:schwarz}
\end{equation}
where $\Fbot=K_0 |dT_0/dz|$ is the imposed bottom flux. For a
given gravity $g$ and conductivity hollow $K_0(T_0)$, convection will
therefore develop for a large enough bottom flux satisfying
Eq.(\ref{eq:schwarz}).

\begin{figure}[htbp]
 \centering
 \includegraphics[width=8cm]{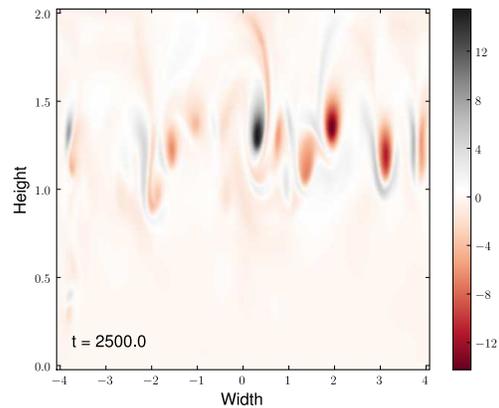}
 \caption{Snapshot of the vorticity field $\na \times\vec{v}$
in a 2-D model with convection and $\kappa$-mechanism.}
 \label{fig:vort}
\end{figure}

Figure \ref{fig:vort} displays a snapshot of the vorticity field obtained
in such a convective simulation at time $t=2500$. In order to ensure that
the thermal relaxation is well achieved, we ran this simulation for a much
longer time than in the purely-radiative case
and the resolution is $256\times256$ gridpoints. As expected, convection
develops
in the middle of the box where the conductivity hollow is
located\footnote{An animation of this simulation is provided at
\url{http://www.ast.obs-mip.fr/users/tgastine/conv_kappa.avi}.}.
Moreover, because of the density contrast across the convection zone,
the vorticity is trapped in strong convective plume downdrafts that
easily overshoot in the stable zone below due to their low P\'eclet number
\citep{Dintrans09}.

\subsection{Evolution of acoustic modes with convection}

To address the time evolution of the acoustic modes that are present
in the DNS with convection, we project as before the physical fields
onto an appropriate acoustic subspace to get the projection coefficient
$c_{\ell n}(t)$. Here $\ell$ denotes the mode degree while $n$ is its
order (with $k_x = (2\pi/L_x)\ell$ and $L_x$ is the width of the box).

\begin{figure}[htbp]
 \centering
 \includegraphics[width=8cm]{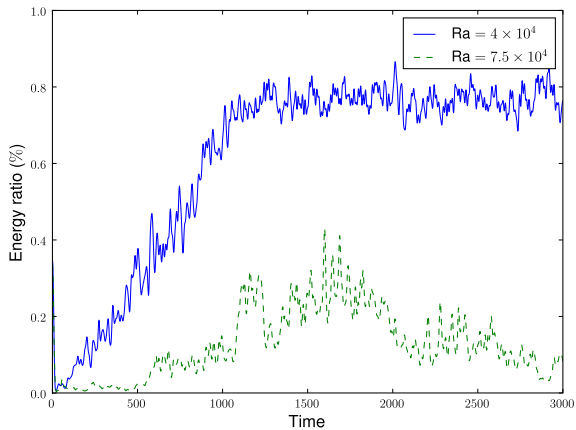}
 \caption{Evolution of the acoustic kinetic energy ratio for two simulations 
  with a weak convection (solid blue
  line) and a stronger one (dashed green
  line).}
 \label{fig:ratio}
\end{figure}

Fig.~\ref{fig:ratio} shows the total energy embedded in acoustic modes
for two simulations with convection that differ by their
Rayleigh number given by

\begin{equation}
 \Ra = -\dfrac{g d_{\text{conv}}^4}{\nu \chi c_p}\dfrac{ds}{dz} =
 \dfrac{(\nabla-\gradad)g d_{\text{conv}}^4}{\nu \chi H_p},
\end{equation}
where $H_p=-(d\ln p_0/dz)^{-1}$ is the pressure scale height,
$d_{\text{conv}}$ the vertical extent of the convection zone,
$\chi=K_0/\rho_0 c_p$ the radiative diffusivity and $s$ the entropy.
This number allows to quantify the strength of the convective
motions, that is, the critical Rayleigh number above which convection
develops is about $10^3$ for polytropic stratifications \cite[e.g.][]{Gough76}.

In the first simulation ($\Ra=4\times 10^4$, solid blue line), the
Rayleigh number is slightly super-critical and the convection intensity
is weak. In this case, we see that the acoustic energy ratio remains
large (i.e. $\gtrsim 80\%$, the remaining $20\%$ being in the convection)
and the nonlinear saturation is reached in the same way than in the
previous radiative models. In other words, the mode propagation is
not affected by convective motions and the convection can be seen as
\emph{frozen-in}.

On the contrary, increasing the Rayleigh number leads to a more vigorous
convection and the acoustic energy ratio is much more affected by
convective plumes (dashed green line). Indeed, after a long transient
during which it reaches a non-trifling value ($\sim 30\%$), it
progressively tends to zero at long times while the convective motions are
responsible for the whole kinetic energy content. The radial oscillations
excited by $\kappa$-mechanism are thus quenched by convection and this
situation is relevant to the red edge where the unstable acoustic modes are 
known to be damped by convective motions occuring at the star's surface.

We thus have shown that the Rayleigh number is a key parameter for this
problem. But further investigations of the convection-pulsation coupling
are in progress, by especially checking time-dependent theories of
convection for which several free parameters can be constrained by such
kind of direct numerical simulations \cite[e.g.][]{YKB}.

\acknowledgments
Calculations were carried out on the CALMIP machine of the University
of Toulouse (\url{http://www.calmip.cict.fr/}).

\makeatletter
\let\clear@thebibliography@page=\relax
\makeatother
\bibliographystyle{Spr-mp-nameyear-cnd}  

\begin{thebibliography}{17}
\ifx \bisbn   \undefined \def \bisbn  #1{ISBN #1}\fi
\ifx \binits  \undefined \def \binits#1{#1} \fi
\ifx \bauthor  \undefined \def \bauthor#1{#1} \fi
\ifx \batitle  \undefined \def \batitle#1{#1} \fi
\ifx \bjtitle  \undefined \def \bjtitle#1{#1}\fi
\ifx \bvolume  \undefined \def \bvolume#1{\textbf{#1}}\fi
\ifx \byear  \undefined \def \byear#1{#1} \fi
\ifx \bissue  \undefined \def \bissue#1{#1} \fi
\ifx \bfpage  \undefined \def \bfpage#1{#1} \fi
\ifx \blpage  \undefined \def \blpage #1{#1} \fi
\ifx \burl  \undefined \def \burl#1{\textsf{#1}} \fi
\ifx \doiurl  \undefined \def \doiurl#1{\textsf{#1}} \fi
\ifx \betal  \undefined \def \betal{\textit{et al.}} \fi
\ifx \binstitute  \undefined \def \binstitute#1{#1} \fi
\ifx \bctitle  \undefined \def \bctitle#1{#1} \fi
\ifx \beditor  \undefined \def \beditor#1{#1} \fi
\ifx \bpublisher  \undefined \def \bpublisher#1{#1} \fi
\ifx \bbtitle  \undefined \def \bbtitle#1{#1} \fi
\ifx \bedition  \undefined \def \bedition#1{#1} \fi
\ifx \bseriesno  \undefined \def \bseriesno#1{#1} \fi
\ifx \blocation  \undefined \def \blocation#1{#1} \fi
\ifx \bsertitle  \undefined \def \bsertitle#1{#1} \fi
\ifx \bsnm \undefined \def \bsnm#1{#1} \fi
\ifx \bsuffix \undefined \def \bsuffix#1{#1} \fi
\ifx \bparticle \undefined \def \bparticle#1{#1} \fi
\ifx \barticle \undefined \def \barticle#1{#1} \fi
\ifx \botherref \undefined \def \botherref #1{#1} \fi
\ifx \url \undefined \def \url#1{\textsf{#1}} \fi
\ifx \bchapter \undefined \def \bchapter#1{#1} \fi
\ifx \bbook \undefined \def \bbook#1{#1} \fi
\ifx \bcomment \undefined \def \bcomment#1{#1} \fi
\ifx \oauthor \undefined \def \oauthor#1{#1} \fi
\ifx \citeauthoryear \undefined \def \citeauthoryear#1{#1} \fi
\def \endbibitem {}

\bibitem[\protect\citeauthoryear{{Baker} and {Kippenhahn}}{1965}]{Baker65}
\begin{barticle}
\bauthor{\bsnm{{Baker}}, \binits{N.}}, \bauthor{\bsnm{{Kippenhahn}},
  \binits{R.}}:
\bjtitle{\apj}
\bvolume{142},
\bfpage{868}
(\byear{1965})
\end{barticle}
\endbibitem

\bibitem[\protect\citeauthoryear{{Baker}}{1987}]{Baker}
\begin{botherref}
\oauthor{\bsnm{{Baker}}, \binits{N.H.}}:
In: {Hillebrandt}, W., {Meyer-Hofmeister}, E., {Thomas}, H.C. (eds.)
Physical Processes in Comets, Stars and Active Galaxies,
p. 105
(1987)
\end{botherref}
\endbibitem

\bibitem[\protect\citeauthoryear{{Bogdan}, {Cattaneo}, and
  {Malagoli}}{1993}]{Bogdan}
\begin{barticle}
\bauthor{\bsnm{{Bogdan}}, \binits{T.J.}}, \bauthor{\bsnm{{Cattaneo}},
  \binits{F.}}, \bauthor{\bsnm{{Malagoli}}, \binits{A.}}:
\bjtitle{\apj}
\bvolume{407},
\bfpage{316}
(\byear{1993})
\end{barticle}
\endbibitem

\bibitem[\protect\citeauthoryear{{Bono}, {Marconi}, and
  {Stellingwerf}}{1999}]{bono99}
\begin{barticle}
\bauthor{\bsnm{{Bono}}, \binits{G.}}, \bauthor{\bsnm{{Marconi}}, \binits{M.}},
  \bauthor{\bsnm{{Stellingwerf}}, \binits{R.F.}}:
\bjtitle{\apjs}
\bvolume{122},
\bfpage{167}
(\byear{1999})
\end{barticle}
\endbibitem

\bibitem[\protect\citeauthoryear{{Chandrasekhar}}{1961}]{Chandra61}
\begin{bbook}
\bauthor{\bsnm{{Chandrasekhar}}, \binits{S.}}:
\bbtitle{{Hydrodynamic and hydromagnetic stability}}.
\bpublisher{Oxford University Press: Clarendon},
\blocation{Oxford}
(\byear{1961})
\end{bbook}
\endbibitem

\bibitem[\protect\citeauthoryear{{Cox}}{1980}]{Cox80}
\begin{bbook}
\bauthor{\bsnm{{Cox}}, \binits{J.P.}}:
\bbtitle{{Theory of stellar pulsation}}.
\bpublisher{Princeton University Press},
\blocation{Princeton}
(\byear{1980})
\end{bbook}
\endbibitem

\bibitem[\protect\citeauthoryear{{Cox} and {Whitney}}{1958}]{cox58}
\begin{barticle}
\bauthor{\bsnm{{Cox}}, \binits{J.P.}}, \bauthor{\bsnm{{Whitney}}, \binits{C.}}:
\bjtitle{\apj}
\bvolume{127},
\bfpage{561}
(\byear{1958})
\end{barticle}
\endbibitem

\bibitem[\protect\citeauthoryear{{Dintrans}}{2009}]{Dintrans09}
\begin{barticle}
\bauthor{\bsnm{{Dintrans}}, \binits{B.}}:
\bjtitle{Communications in Asteroseismology}
\bvolume{158},
\bfpage{45}
(\byear{2009})
\end{barticle}
\endbibitem

\bibitem[\protect\citeauthoryear{{Dintrans} and
  {Brandenburg}}{2004}]{dintrans04}
\begin{barticle}
\bauthor{\bsnm{{Dintrans}}, \binits{B.}}, \bauthor{\bsnm{{Brandenburg}},
  \binits{A.}}:
\bjtitle{A\&A}
\bvolume{421},
\bfpage{775}
(\byear{2004})
\end{barticle}
\endbibitem

\bibitem[\protect\citeauthoryear{{Eddington}}{1917}]{Eddington1917}
\begin{barticle}
\bauthor{\bsnm{{Eddington}}, \binits{A.S.}}:
\bjtitle{The Observatory}
\bvolume{40},
\bfpage{290}
(\byear{1917})
\end{barticle}
\endbibitem

\bibitem[\protect\citeauthoryear{{Gastine} and {Dintrans}}{2008a}]{paperI}
\begin{barticle}
\bauthor{\bsnm{{Gastine}}, \binits{T.}}, \bauthor{\bsnm{{Dintrans}},
  \binits{B.}}:
\bjtitle{\aap}
\bvolume{484},
\bfpage{29}
(\byear{2008}a)
\end{barticle}
\endbibitem

\bibitem[\protect\citeauthoryear{{Gastine} and {Dintrans}}{2008b}]{paperII}
\begin{barticle}
\bauthor{\bsnm{{Gastine}}, \binits{T.}}, \bauthor{\bsnm{{Dintrans}},
  \binits{B.}}:
\bjtitle{\aap}
\bvolume{490},
\bfpage{743}
(\byear{2008}b)
\end{barticle}
\endbibitem

\bibitem[\protect\citeauthoryear{{Gough} \textit{et~al.}}{1976}]{Gough76}
\begin{barticle}
\bauthor{\bsnm{{Gough}}, \binits{D.O.}}, \bauthor{\bsnm{{Moore}},
  \binits{D.R.}}, \bauthor{\bsnm{{Spiegel}}, \binits{E.A.}},
  \bauthor{\bsnm{{Weiss}}, \binits{N.O.}}:
\bjtitle{\apj}
\bvolume{206},
\bfpage{536}
(\byear{1976})
\end{barticle}
\endbibitem

\bibitem[\protect\citeauthoryear{{Unno} \textit{et~al.}}{1989}]{Unno-book}
\begin{bbook}
\bauthor{\bsnm{{Unno}}, \binits{W.}}, \bauthor{\bsnm{{Osaki}}, \binits{Y.}},
  \bauthor{\bsnm{{Ando}}, \binits{H.}}, \bauthor{\bsnm{{Saio}}, \binits{H.}},
  \bauthor{\bsnm{{Shibahashi}}, \binits{H.}}:
\bbtitle{{Nonradial oscillations of stars}}.
\bpublisher{University of Tokyo Press},
\blocation{Tokyo}
(\byear{1989})
\end{bbook}
\endbibitem

\bibitem[\protect\citeauthoryear{{Valdettaro} \textit{et~al.}}{2007}]{LSB}
\begin{barticle}
\bauthor{\bsnm{{Valdettaro}}, \binits{L.}}, \bauthor{\bsnm{{Rieutord}},
  \binits{M.}}, \bauthor{\bsnm{{Braconnier}}, \binits{T.}},
  \bauthor{\bsnm{{Fraysse}}, \binits{V.}}:
\bjtitle{JCoAM}
\bvolume{205}(\bissue{1}),
\bfpage{382}
(\byear{2007})
\end{barticle}
\endbibitem

\bibitem[\protect\citeauthoryear{{Yecko}, {Koll{\'a}th}, and
  {Buchler}}{1998}]{YKB}
\begin{barticle}
\bauthor{\bsnm{{Yecko}}, \binits{P.A.}}, \bauthor{\bsnm{{Koll{\'a}th}},
  \binits{Z.}}, \bauthor{\bsnm{{Buchler}}, \binits{J.R.}}:
\bjtitle{\aap}
\bvolume{336},
\bfpage{553}
(\byear{1998})
\end{barticle}
\endbibitem

\bibitem[\protect\citeauthoryear{{Zhevakin}}{1963}]{zhevakin}
\begin{barticle}
\bauthor{\bsnm{{Zhevakin}}, \binits{S.A.}}:
\bjtitle{\araa}
\bvolume{1},
\bfpage{367}
(\byear{1963})
\end{barticle}
\endbibitem

\end{thebibliography}

\end{document}